\documentclass[twocolumn,twocolappendix]{aastex63}
\usepackage[utf8]{inputenc}
\usepackage{placeins}
\usepackage{graphicx}
\usepackage{tabularx}
\usepackage{amsmath}
\usepackage{physics}

\usepackage{dcolumn}
\newcolumntype{d}[1]{D{.}{.}{#1}}

\newcommand{\alos}{$a_\mathrm{los}$ }

\begin{document}

\title{The Anomalous Acceleration of PSR J2043+1711: \\ Long-Period Orbital Companion or Stellar Flyby?}

\author[0000-0002-7746-8993]{Thomas Donlon II}
\affiliation{Department of Physics and Astronomy, University of Alabama in Huntsville, 301 North Sparkman Drive, Huntsville, AL 35816, USA}
\correspondingauthor{Thomas Donlon II}
\email{thomas.donlon@uah.edu}

\author[0000-0001-6711-8140]{Sukanya Chakrabarti}
\affiliation{Department of Physics and Astronomy, University of Alabama in Huntsville, 301 North Sparkman Drive, Huntsville, AL 35816, USA}

\author[0000-0003-0721-651X]{Michael T. Lam}
\affiliation{SETI Institute, 339 N Bernardo Ave Suite 200, Mountain View, CA 94043, USA}
\affiliation{School of Physics and Astronomy, Rochester Institute of Technology, Rochester, NY 14623, USA}
\affiliation{Laboratory for Multiwavelength Astrophysics, Rochester Institute of Technology, Rochester, NY 14623, USA}

\author[0000-0001-8832-4488]{Daniel Huber} 
\affiliation{Institute for Astronomy, University of Hawai`i, 2680 Woodlawn Drive, Honolulu HI 96822}
\affiliation{Sydney Institute for Astronomy (SIfA), School of Physics, University of Sydney, NSW 2006, Australia}

\author[0000-0003-3244-5357]{Daniel Hey} 
\affiliation{Institute for Astronomy, University of Hawai`i, 2680 Woodlawn Drive, Honolulu HI 96822}

\author[0000-0003-2558-3102]{Enrico Ramirez-Ruiz}
\affiliation{Department of Astronomy and Astrophysics, University of California, Santa Cruz, CA 95064, USA}

\author[0000-0003-4631-1149]{Benjamin Shappee} 
\affiliation{Institute for Astronomy, University of Hawai`i, 2680 Woodlawn Drive, Honolulu HI 96822}

\author[0000-0001-6295-2881]{David L. Kaplan}
\affiliation{Center for Gravitation, Cosmology and Astrophysics, Department of Physics, University of Wisconsin-Milwaukee,\\ P.O. Box 413, Milwaukee, WI 53201, USA}


\author[0000-0001-5134-3925]{Gabriella Agazie}
\affiliation{Center for Gravitation, Cosmology and Astrophysics, Department of Physics, University of Wisconsin-Milwaukee,\\ P.O. Box 413, Milwaukee, WI 53201, USA}

\author[0000-0002-8935-9882]{Akash Anumarlapudi}
\affiliation{Center for Gravitation, Cosmology and Astrophysics, Department of Physics, University of Wisconsin-Milwaukee,\\ P.O. Box 413, Milwaukee, WI 53201, USA}

\author[0000-0003-0638-3340]{Anne M. Archibald}
\affiliation{Newcastle University, NE1 7RU, UK}

\author{Zaven Arzoumanian}
\affiliation{X-Ray Astrophysics Laboratory, NASA Goddard Space Flight Center, Code 662, Greenbelt, MD 20771, USA}

\author[0000-0003-2745-753X]{Paul T. Baker}
\affiliation{Department of Physics and Astronomy, Widener University, One University Place, Chester, PA 19013, USA}

\author[0000-0003-3053-6538]{Paul R. Brook}
\affiliation{Institute for Gravitational Wave Astronomy and School of Physics and Astronomy, University of Birmingham, Edgbaston, Birmingham B15 2TT, UK}

\author[0000-0002-6039-692X]{H. Thankful Cromartie}
\affiliation{National Research Council Research Associate, National Academy of Sciences, Washington, DC 20001, USA resident at Naval Research Laboratory, Washington, DC 20375, USA}

\author[0000-0002-1529-5169]{Kathryn Crowter}
\affiliation{Department of Physics and Astronomy, University of British Columbia, 6224 Agricultural Road, Vancouver, BC V6T 1Z1, Canada}

\author[0000-0002-2185-1790]{Megan E. DeCesar}
\affiliation{George Mason University, Fairfax, VA 22030, resident at the U.S. Naval Research Laboratory, Washington, DC 20375, USA}

\author[0000-0002-6664-965X]{Paul B. Demorest}
\affiliation{National Radio Astronomy Observatory, 1003 Lopezville Rd., Socorro, NM 87801, USA}

\author[0000-0001-8885-6388]{Timothy Dolch}
\affiliation{Department of Physics, Hillsdale College, 33 E. College Street, Hillsdale, MI 49242, USA}
\affiliation{Eureka Scientific, 2452 Delmer Street, Suite 100, Oakland, CA 94602-3017, USA}

\author[0000-0001-7828-7708]{Elizabeth C. Ferrara}
\affiliation{Department of Astronomy, University of Maryland, College Park, MD 20742, USA}
\affiliation{Center for Research and Exploration in Space Science and Technology, NASA/GSFC, Greenbelt, MD 20771}
\affiliation{NASA Goddard Space Flight Center, Greenbelt, MD 20771, USA}

\author[0000-0001-5645-5336]{William Fiore}
\affiliation{Department of Physics and Astronomy, West Virginia University, P.O. Box 6315, Morgantown, WV 26506, USA}
\affiliation{Center for Gravitational Waves and Cosmology, West Virginia University, Chestnut Ridge Research Building, Morgantown, WV 26505, USA}

\author[0000-0001-8384-5049]{Emmanuel Fonseca}
\affiliation{Department of Physics and Astronomy, West Virginia University, P.O. Box 6315, Morgantown, WV 26506, USA}
\affiliation{Center for Gravitational Waves and Cosmology, West Virginia University, Chestnut Ridge Research Building, Morgantown, WV 26505, USA}

\author[0000-0001-7624-4616]{Gabriel E. Freedman}
\affiliation{Center for Gravitation, Cosmology and Astrophysics, Department of Physics, University of Wisconsin-Milwaukee,\\ P.O. Box 413, Milwaukee, WI 53201, USA}

\author[0000-0001-6166-9646]{Nate Garver-Daniels}
\affiliation{Department of Physics and Astronomy, West Virginia University, P.O. Box 6315, Morgantown, WV 26506, USA}
\affiliation{Center for Gravitational Waves and Cosmology, West Virginia University, Chestnut Ridge Research Building, Morgantown, WV 26505, USA}

\author[0000-0001-8158-683X]{Peter A. Gentile}
\affiliation{Department of Physics and Astronomy, West Virginia University, P.O. Box 6315, Morgantown, WV 26506, USA}
\affiliation{Center for Gravitational Waves and Cosmology, West Virginia University, Chestnut Ridge Research Building, Morgantown, WV 26505, USA}

\author[0000-0003-4090-9780]{Joseph Glaser}
\affiliation{Department of Physics and Astronomy, West Virginia University, P.O. Box 6315, Morgantown, WV 26506, USA}
\affiliation{Center for Gravitational Waves and Cosmology, West Virginia University, Chestnut Ridge Research Building, Morgantown, WV 26505, USA}

\author[0000-0003-1884-348X]{Deborah C. Good}
\affiliation{Department of Physics and Astronomy, University of Montana, 32 Campus Drive, Missoula, MT 59812}

\author[0000-0003-2742-3321]{Jeffrey S. Hazboun}
\affiliation{Department of Physics, Oregon State University, Corvallis, OR 97331, USA}

\author[0000-0003-1059-9603]{Mark Huber} 
\affiliation{Institute for Astronomy, University of Hawai`i, 2680 Woodlawn Drive, Honolulu HI 96822}

\author[0000-0003-1082-2342]{Ross J. Jennings}
\altaffiliation{NANOGrav Physics Frontiers Center Postdoctoral Fellow}
\affiliation{Department of Physics and Astronomy, West Virginia University, P.O. Box 6315, Morgantown, WV 26506, USA}
\affiliation{Center for Gravitational Waves and Cosmology, West Virginia University, Chestnut Ridge Research Building, Morgantown, WV 26505, USA}

\author[0000-0001-6607-3710]{Megan L. Jones}
\affiliation{Center for Gravitation, Cosmology and Astrophysics, Department of Physics, University of Wisconsin-Milwaukee,\\ P.O. Box 413, Milwaukee, WI 53201, USA}

\author[0000-0002-0893-4073]{Matthew Kerr}
\affiliation{Space Science Division, Naval Research Laboratory, Washington, DC 20375-5352, USA}

\author[0000-0003-1301-966X]{Duncan R. Lorimer}
\affiliation{Department of Physics and Astronomy, West Virginia University, P.O. Box 6315, Morgantown, WV 26506, USA}
\affiliation{Center for Gravitational Waves and Cosmology, West Virginia University, Chestnut Ridge Research Building, Morgantown, WV 26505, USA}

\author[0000-0001-5373-5914]{Jing Luo}
\altaffiliation{Deceased}
\affiliation{Department of Astronomy \& Astrophysics, University of Toronto, 50 Saint George Street, Toronto, ON M5S 3H4, Canada}

\author[0000-0001-5229-7430]{Ryan S. Lynch}
\affiliation{Green Bank Observatory, P.O. Box 2, Green Bank, WV 24944, USA}

\author[0000-0001-5481-7559]{Alexander McEwen}
\affiliation{Center for Gravitation, Cosmology and Astrophysics, Department of Physics, University of Wisconsin-Milwaukee,\\ P.O. Box 413, Milwaukee, WI 53201, USA}

\author[0000-0001-7697-7422]{Maura A. McLaughlin}
\affiliation{Department of Physics and Astronomy, West Virginia University, P.O. Box 6315, Morgantown, WV 26506, USA}
\affiliation{Center for Gravitational Waves and Cosmology, West Virginia University, Chestnut Ridge Research Building, Morgantown, WV 26505, USA}

\author[0000-0002-4642-1260]{Natasha McMann}
\affiliation{Department of Physics and Astronomy, Vanderbilt University, 2301 Vanderbilt Place, Nashville, TN 37235, USA}

\author[0000-0001-8845-1225]{Bradley W. Meyers}
\affiliation{Department of Physics and Astronomy, University of British Columbia, 6224 Agricultural Road, Vancouver, BC V6T 1Z1, Canada}
\affiliation{International Centre for Radio Astronomy Research, Curtin University, Bentley, WA 6102, Australia}

\author[0000-0002-3616-5160]{Cherry Ng}
\affiliation{Dunlap Institute for Astronomy and Astrophysics, University of Toronto, 50 St. George St., Toronto, ON M5S 3H4, Canada}

\author[0000-0002-6709-2566]{David J. Nice}
\affiliation{Department of Physics, Lafayette College, Easton, PA 18042, USA}

\author[0000-0001-5465-2889]{Timothy T. Pennucci}
\affiliation{Institute of Physics and Astronomy, E\"{o}tv\"{o}s Lor\'{a}nd University, P\'{a}zm\'{a}ny P. s. 1/A, 1117 Budapest, Hungary}

\author[0000-0002-8509-5947]{Benetge B. P. Perera}
\affiliation{Arecibo Observatory, HC3 Box 53995, Arecibo, PR 00612, USA}

\author[0000-0002-8826-1285]{Nihan S. Pol}
\affiliation{Department of Physics and Astronomy, Vanderbilt University, 2301 Vanderbilt Place, Nashville, TN 37235, USA}

\author[0000-0002-2074-4360]{Henri A. Radovan}
\affiliation{Department of Physics, University of Puerto Rico, Mayag\"{u}ez, PR 00681, USA}

\author[0000-0001-5799-9714]{Scott M. Ransom}
\affiliation{National Radio Astronomy Observatory, 520 Edgemont Road, Charlottesville, VA 22903, USA}

\author[0000-0002-5297-5278]{Paul S. Ray}
\affiliation{Space Science Division, Naval Research Laboratory, Washington, DC 20375-5352, USA}

\author[0000-0003-4391-936X]{Ann Schmiedekamp}
\affiliation{Department of Physics, Penn State Abington, Abington, PA 19001, USA}

\author[0000-0002-1283-2184]{Carl Schmiedekamp}
\affiliation{Department of Physics, Penn State Abington, Abington, PA 19001, USA}

\author[0000-0002-7283-1124]{Brent J. Shapiro-Albert}
\affiliation{Department of Physics and Astronomy, West Virginia University, P.O. Box 6315, Morgantown, WV 26506, USA}
\affiliation{Center for Gravitational Waves and Cosmology, West Virginia University, Chestnut Ridge Research Building, Morgantown, WV 26505, USA}
\affiliation{Giant Army, 915A 17th Ave, Seattle WA 98122}

\author[0000-0001-9784-8670]{Ingrid H. Stairs}
\affiliation{Department of Physics and Astronomy, University of British Columbia, 6224 Agricultural Road, Vancouver, BC V6T 1Z1, Canada}

\author[0000-0002-7261-594X]{Kevin Stovall}
\affiliation{National Radio Astronomy Observatory, 1003 Lopezville Rd., Socorro, NM 87801, USA}

\author[0000-0002-2820-0931]{Abhimanyu Susobhanan}
\affiliation{Max-Planck-Institut f\"{u}r Gravitationsphysik (Albert-Einstein-Institut), Callinstrasse 38, D-30167, Hannover, Germany}

\author[0000-0002-1075-3837]{Joseph K. Swiggum}
\altaffiliation{NANOGrav Physics Frontiers Center Postdoctoral Fellow}
\affiliation{Department of Physics, Lafayette College, Easton, PA 18042, USA}

\author[0000-0002-2471-8442]{Michael A. Tucker} 
\altaffiliation{CCAPP Fellow}
\affiliation{Center for Cosmology and AstroParticle Physics, 191 W Woodruff Ave, Columbus, OH 43210}
\affiliation{Department of Astronomy, The Ohio State University, 140 W 18th Ave, Columbus, OH 43210}

\author[0000-0001-9678-0299]{Haley M. Wahl}
\affiliation{Department of Physics and Astronomy, West Virginia University, P.O. Box 6315, Morgantown, WV 26506, USA}
\affiliation{Center for Gravitational Waves and Cosmology, West Virginia University, Chestnut Ridge Research Building, Morgantown, WV 26505, USA}


\begin{abstract}
Based on the rate of change of its orbital period, PSR J2043+1711 has a substantial peculiar acceleration of 3.5 $\pm$ 0.8 mm/s/yr, which deviates from the acceleration predicted by equilibrium Milky Way models at a $4\sigma$ level. The magnitude of the peculiar acceleration is too large to be explained by disequilibrium effects of the Milky Way interacting with orbiting dwarf galaxies ($\sim$1 mm/s/yr), and too small to be caused by period variations due to the pulsar being a redback. We identify and examine two plausible causes for the anomalous acceleration: a stellar flyby, and a long-period orbital companion. We identify a main-sequence star in \textit{Gaia} DR3 and Pan-STARRS DR2 with the correct mass, distance, and on-sky position to potentially explain the observed peculiar acceleration. However, the star and the pulsar system have substantially different proper motions, indicating that they are not gravitationally bound. However, it is possible that this is an unrelated star that just happens to be located near J2043+1711 along our line of sight (chance probability of 1.6\%). Therefore, we also constrain possible orbital parameters for a circumbinary companion in a hierarchical triple system with J2043+1711; the changes in the spindown rate of the pulsar are consistent with an outer object that has an orbital period of 60 kyr, a companion mass of 0.3 $M_\odot$ (indicative of a white dwarf or low-mass star), and a semi-major axis of 1900 AU. Continued timing and/or future faint optical observations of J2043+1711 may eventually allow us to differentiate between these scenarios. \\\vspace{0.5cm}
\end{abstract}

\section{Introduction} \label{sec:intro}

Incredibly precise time-series measurements of pulsars have been pushing the boundaries of astrophysics for several decades. Pulsars have been widely used as tests of strong-field general relativity \citep{DamourTaylor1992,Stairs2003,Antoniadis2013,WeisbergHuang2016}, and have recently been used to find the first strong evidence of a gravitational wave background \citep{nanograv_gwb2023,epta_gwb2023,parkes_gwb2023,cpta_gwb2023}. The ability to use pulsars as accelerometers is becoming increasingly relevant, and binary millisecond pulsars have already been used to map our Galaxy's gravitational field without the kinematic assumptions of dynamical equilibrium or symmetry \citep{Chakrabarti2021,Moran2023,Donlon2024}.  The spin period of solitary millisecond pulsars could in principle also be used if the dependence on the magnetic braking were known, but currently this procedure leads to large uncertainties \citep{Phillips2021}.  The variety of relevant uses for pulsars make them extremely important astrophysical tools; as such, understanding anomalies in the properties of individual pulsars is crucial, as the physics of pulsars can have widespread implications across multiple fields. 

PSR J2043+1711 was first discovered by the \textit{Fermi} Large Area Telescope \citep{Atwood2009,Acero2015} as a gamma-ray source, and was then shown to be a millisecond pulsar by \cite{Guillemot2012} using the Nan\c{c}ay Radio Telescope. These follow-up observations showed that the pulsar was in a 1.48 day period orbit around a companion, which was likely a helium white dwarf due to its low mass. Due to the pulsar's particularly stable rotation rate, it was quickly added to the list of pulsars observed by the NANOGrav collaboration, and a measurement of the mass of the pulsar from Shapiro delay was included in the NANOGrav 9-year dataset \citep{nanograv9yr}. The most up-to-date parameters for J2043+1711 use timing data with a 9.2 year baseline from the NANOGrav 15-year data release \citep{nanograv15yr}, where the fit was performed using the ELL1 binary model. 

We show that the observed acceleration of J2043+1711, when computed using the parameters provided by the NANOGrav 15-year data release, is statistically inconsistent with predictions from commonly used models for the gravitational potential of the Galaxy.  We explore several scenarios that could potentially generate this observed deviation, including dynamical disequilibrium in the Galaxy, a stellar flyby, a circumbinary orbital companion, and J2043+1711 being a spider pulsar. The observed acceleration of J2043+1711 is statistically inconsistent with predictions from commonly used Galactic potential models, because the magnitude of accelerations caused by dynamical disequilibria are too small to explain the observed peculiar acceleration. Similarly, if J2043+1711 were a spider pulsar, it would experience an intrinsic acceleration that is orders of magnitude larger than what we observe for this system. This leaves the stellar flyby and circumbinary companion scenarios as the only two viable options that could explain the observed peculiar acceleration of the system. 

Each of these possibilities has distinct ramifications for its respective field: it is now clear that dynamical disequilibrium is essential to understanding our Galaxy, and can be constrained using pulsar timing data \citep[e.g.][]{Antoja2018,PetersenPenarrubia2021,Donlon2024}; stellar flybys are incredibly rare, and have only been observed a handful of times in young and proto-stellar systems \citep{Dai2015,Borchert2022,Cuello2023}; and it is unknown how common circumbinary companions to binary millisecond pulsars are, although at least two such hierarchical triple systems have been shown to exist \citep{Thorsett1993,Thorsett1999,Ransom2014} and others have been proposed \citep[e.g.][]{Nieder2022}. Additionally, constraining the properties of circumbinary objects could prove useful for understanding the formation processes of pulsar systems. J2043+1711 is a uniquely interesting system that allows us to explore each of these ideas. 

This would not be the first time that an unexpected acceleration of a pulsar has led to the discovery of an orbital companion. \cite{Matthews2016} argued that J1024$-$0719, at the time believed to be an isolated pulsar, had anomalous velocity and acceleration measurements that were consistent with an orbital companion at a large distance from the pulsar. Later, \cite{Kaplan2016} and \cite{Bassa2016} simultaneously showed that J1024$-$0719 was in a wide orbit around with a K star. While we are unable to determine the cause of the peculiar acceleration for J2043+1711, we hope that the arguments presented in this work will eventually lead to the confirmation of an additional orbital companion or a stellar flyby. 


\section{Computing An Acceleration}


The NANOGrav 15 year data release incorporated Post-Keplerian orbital parameters such as $\dot{P}_{\rm b}^{\rm Obs}$ based on a statistical significance $F$-test that is roughly equivalent to including any such parameter that has significance of $\sim 3\sigma$ or more; see \cite{nanograv15yr} for more details.  PSR J2043+1711 was found to have a significant $\dot{P}_{\rm b}^{\rm Obs}$ using this criterion.

The observed change in the orbital period ($\dot{P}_b^{\rm Obs}$) of a millisecond pulsar binary system can be decomposed into several independent effects: 

\begin{equation} \label{eq:eq1}
    \dot{P}_b^\mathrm{Obs} = \dot{P}_b^\mathrm{Shk} + \dot{P}_b^\mathrm{GR} + \dot{P}_b^{\rm Int} + \dot{P}_b^\Phi.
\end{equation} The term $\dot{P}_b^\mathrm{Shk}$ is the change in the orbital period ($P_b$) due to the Shklovskii Effect \citep{Shklovskii1970}, which is caused by transverse motion of the source on the sky leading to an apparent change in the orbital period. The term $\dot{P}_b^\mathrm{GR}$ is the amount that $P_b$ decreases due to the relativistic decay of the binary orbit from the emission of gravitational waves \citep{PetersMathews1963,WeisbergHuang2016}, and is a function of the orbital period, eccentricity and masses of the objects in the system. J2043+1711 is not a particularly relativistic system, so in this case $\dot{P}_b^\mathrm{GR}$ is an order of magnitude smaller than the other terms. Errors were propagated in the standard way for these terms based on individual uncertainties in distance, proper motion, the respective masses, etc. 

The term $\dot{P}_b^{\rm Int}$ is a catch-all term for various interactions between a pulsar and its orbital companion. This includes tidal effects, radiative effects, exchanges of mass and/or angular momentum, etc. These effects can be caused either by the strong emission jets radiated by the pulsar interacting with the companion, stellar evolution of the companion causing a change in the configuration of the orbit, or some other complicated process.

The remaining term, $\dot{P}_b^\Phi$, corresponds to the amount that $P_b$ changes due to the gravitational potential at the position of the pulsar. The line-of-sight acceleration of the pulsar (relative to the Solar system barycenter) can be calculated as

\begin{equation} \label{eq:eq2}
    a_\mathrm{los} = \mathbf{a} \cdot \hat{\mathbf{x}}_\mathrm{los} = \frac{\dot{P}_b^\mathrm{\Phi}}{P_b}c,
\end{equation} where $\hat{\mathbf{x}}_\mathrm{los}$ is the unit vector pointing from the Sun to the pulsar. Note that this is not an absolute acceleration with respect to the inertial frame of the Galaxy; rather it is the difference between the potential gradient at the location of the pulsar and the potential gradient at the Solar position. As such, it is not affected by uncertainties in the Solar location or velocities.

The term $\dot{P}_b^\Phi$ is also commonly written as $\dot{P}_b^\mathrm{Gal}$, as the relevant quantity is typically the acceleration due to the large-scale gravitational field of the Milky Way (MW), while other effects that can cause accelerations are presumed to be negligible. However, in this case it is helpful to split the line-of-sight acceleration into two distinct components; \begin{equation} \label{eq:acc_breakdown}
    a_\mathrm{los} = a_\mathrm{los}^\mathrm{Gal} + a_\mathrm{los}^\mathrm{Pec},
\end{equation} where $a_\mathrm{los}^\mathrm{Gal}$ is the line-of-sight acceleration due to the large-scale (smooth) gravitational field of the MW, and $a_\mathrm{los}^\mathrm{Pec}$ is the \textit{peculiar} line-of-sight acceleration of the pulsar due to additional effects. 

\begin{table}[]
    \centering
    \begin{tabularx}{0.48\textwidth}{@{\extracolsep{\fill}} lr} \hline \hline
        \multicolumn{2}{c}{PSR J2043+1711} \\ \hline
        \multicolumn{2}{c}{\textit{Fit Parameters}} \\ \hline
        Spin Frequency Epoch & MJD 57413.000000000 \\
        Binary Epoch & TASC = MJD 57413.501338113(3) \\
        $f$ & 420.18944316950783(10) s$^{-1}$ \\
        $f^{(1)}$ & $-$9.25932(2)E$-$16 s$^{-2}$ \\
        $f^{(2)*}$ & 3.5(24)E$-$27 s$^{-3}$ \\
        $f^{(3)*}$ & 5.8(27)E$-$34 s$^{-4}$ \\
        $P_b$ & 1.482290786388(6) days \\
        $\dot{P}_b^\mathrm{Obs}$ & 1.02(12)E$-$13 s/s \\
        $\varpi$ & 0.64(4) mas \\
        $\sin i$ & 0.990(1) \\ \hline
        \multicolumn{2}{c}{\textit{Derived Parameters}} \\ \hline
        $d$ & 1.56(10) kpc \\
        $\mu_\alpha$ & $-$5.703(11) mas yr$^{-1}$\\
        $\mu_\delta$ & $-$10.841(17) mas yr$^{-1}$\\
        $e$ & 5.01(8)E$-$6 \\ 
        $M_p$ & 1.62(10) $M_\odot$ \\
        $M_c$ & 0.190(7) $M_\odot$ \\
        $\dot{P}_b^\mathrm{Shk}$ & 7.3(5)E$-$14 s/s \\ 
        $\dot{P}_b^\mathrm{GR}$ & $-$2.86(16)E$-$15 s/s \\ \hline \hline
        \multicolumn{2}{c}{\textit{Gaia} DR3 1811439569904158208} \\ \hline
        $m_G$ & 17.27 \\
        $M_G$ & 6.0$^{+0.3}_{-0.2}$ \\
        $B_P - R_P$ & 1.13 \\
        $\mu_\alpha$ & 3.81(7) mas yr$^{-1}$\\
        $\mu_\delta$ & 0.13(6) mas yr$^{-1}$\\
        $\varpi$ & 0.57(7) mas \\
        RUWE & 0.93 \\
        $M_\mathrm{FLAME}$ & 0.80$^{+0.04}_{-0.05}$ $M_\odot$ \\
        $g_\mathrm{PS}$ & 17.912(4) \\
        $r_\mathrm{PS}$ & 17.341(4) \\
        $i_\mathrm{PS}$ & 16.935(3) \\
        $z_\mathrm{PS}$ & 16.850(3) \\
        $y_\mathrm{PS}$ & 16.946(7) \\
        RV$^*$ & $-$4(50) km/s \\\hline
    \end{tabularx}
    \caption{Properties of PSR J2043+1711 and \textit{Gaia} DR3 1811439569904158208. The quantities marked with * are fit separately and are not part of the NANOGrav 15-year data set or \textit{Gaia} DR3 catalog. }
    \label{tab:properties}
\end{table}

We are able to calculate \alos for PSR J2043+1711 using Equations \ref{eq:eq1} and \ref{eq:eq2} with the timing solution from the NANOGrav 15-year data set \citep[][see Table \ref{tab:properties}]{nanograv15yr}, yielding \alos = 2.2 $\pm$ 0.8 mm/s/yr.  Using the MilkyWayPotential2022 model from the \textbf{Gala} package \citep{Gala}, we estimate the contribution from the smooth Galactic potential to be $a_\mathrm{los}^\mathrm{Gal} = -1.47 \pm 0.10$ mm/s/yr, where the uncertainty in this value arises from the uncertainty in the distance to the pulsar. This potential model is fit to observed kinematic data, and therefore represents a reasonable time-static approximation to the true Galactic potential. Using other common kinematic potential models only changes the value of $a_\mathrm{los}^\mathrm{Gal}$ by a few percent, which does not change the following analysis. 

This leads to a peculiar line-of-sight acceleration of $a_\mathrm{los}^\mathrm{Pec} = 3.7 \pm 0.8$ mm/s/yr. However, \cite{Donlon2024} showed that pulsar accelerations have a global trend relative to kinematic MW models; this effect might be related to dynamical disequilibrium features or a peculiar Solar acceleration, but cannot be explained by processes intrinsic to pulsar timing, and as a result contributes a bias to measured accelerations. The amount of bias expected at the position of J2043+1711 is 0.5$\pm$0.1 mm/s/yr, which was determined using the procedure in Section III of \cite{Donlon2024} (see Appendix \ref{app:d24_bias} for more information on how this calculation is done). Subtracting this amount to remove the bias reduces the peculiar line-of-sight acceleration to $a_\mathrm{los}^\mathrm{Pec} = 3.2 \pm 0.8$ mm/s/yr. This is roughly double the strength of the expected Galactic acceleration, and represents a 4$\sigma$ deviation from the underlying Galactic potential models. The possible causes for this peculiar acceleration are explored in the following sections.

Note that we report symmetrical uncertainties throughout this work, although this may not be the case for individual quantities -- for example, distance has an asymmetric uncertainty distribution. \cite{Donlon2024} show in their Figure 2 and related discussion that this treatment is appropriate for most pulsars (including J2043+1711), and produces very similar values and uncertainties for inferred accelerations. We have confirmed that this assumption is appropriate for the various components of $\dot{P}_b$, which have only slightly non-Gaussian posteriors.

This process was also carried out for all other pulsars in the \cite{Donlon2024} dataset, but only J2043+1711 was found to have a peculiar acceleration that deviates from the expected Galactic acceleration by more than $3\sigma$. This further suggests that the anomalous acceleration of J2043+1711 is actually due to some process occurring specifically for that pulsar -- if this phenomenon was a result of incorrectly calculating accelerations for pulsars in general, then we would expect to see the same type of anomalous accelerations in many pulsars, which is not the case.

\section{Is J2043+1711 a Spider Pulsar?}

\begin{table}[]
    \centering
    \begin{tabular}{lr} \hline \hline
        PSR & $\dot{P}_b/P_b$ \\
         & \textit{(1/s)} \\ \hline
        \textbf{J2043+1711} & \textbf{7.8E$-$19} \\
        J0024$-$7204W & $-$1.5E$-$15 \\
        J1023+0038 & $-$4.3E$-$15 \\	
        J1048+2339 & $-$1.2E$-$14 \\
        J1227$-$4853 & $-$3.5E$-$14 \\
        J1622$-$0315	& 2.2E$-$15 \\
        J1717+4308A	& 9.5E$-$16 \\
        J1723$-$2837	& $-$6.6E$-$14	\\	
        J1740$-$5340A & 2.6E$-$17 \\
        J1748$-$2446ad & $-$2.4E$-$17 \\
        J1803$-$6707 & $-$3.5E$-$15 \\
        J2039$-$5617	& 4.1E$-$16	\\
        J2215+5135	& $-$3.0E$-$14	\\
        J2339$-$0533	& $-$1.2E$-$14 \\ \hline
    \end{tabular}
    \caption{Fractional change in orbital period over time for J2043+1711 compared to those of redback pulsars.}
    \label{tab:redbacks}
\end{table}

The first term in Equation \ref{eq:eq1} that we wish to consider is $\dot{P}_b^{\rm Int}$, because it is plausible that what the observed peculiar acceleration is not actually a physical acceleration at all, but a variation in the orbital period of the J2043+1711 system due to some sort of complicated interaction between the bound objects. One common type of system that could experience a large $\dot{P}_b^{\rm Int}$ is a redback pulsar.  

Spider pulsars are systems where a star has transferred mass to a neutron star companion, recycling the pulsar \citep{Roberts2013}. If the system is configured so that the pulsar wind interacts with the donor star, this can ablate material away from the companion, leading to evaporation of the donor star on Gyr timescales. The two major categories of spider pulsars are black widows ($M_c < 0.1$ M$_\odot$) and redbacks ($M_c > 0.1$ M$_\odot$); in redbacks, the donor star usually becomes a low-mass but overly-luminous star with an extended size and a seemingly normal spectral type \citep{DeVito2020}. 

The orbital period, companion mass, and spin period of the J2043+1711 system can potentially classify the binary as a redback \citep{Swihart2022}, although J2043+1711 is located in an overlapping region shared by redbacks and millisecond pulsar/white dwarf systems. Spider pulsars can experience intrinsic changes in their orbital periods and often have multiple orbital period derivatives, presumably due to tidal interactions with their extended companion and the evaporated outflow from the donor star; if J2043+1711 is in fact a redback, this could plausibly explain the observed peculiar acceleration of the system without requiring any additional nearby objects. 

\subsection{Eclipses}

One way to determine whether a pulsar is a spider is through eclipses; if a pulsar is eclipsed by its companion, then its radio jets are occulted by the evaporated material, confirming that the system is a spider. J2043+1711 does not have apparent eclipses in the phase-wrapped residual of the NANOGrav 15-year dataset. This does not positively rule out the case that J2043+1711 is a redback, as there are many non-eclipsing systems that are confirmed to be spiders \citep[i.e.][which discusses 3 non-eclipsing spiders]{nanograv15yr}; however, it is a point that argues against the redback scenario, as a significant Shapiro delay and an inclination angle of $i\sim83^\circ$ has been measured for the J2043+1711 system, which makes it likely that we would detect eclipses if the companion was being strongly irradiated. It is worth noting, though, that because the system would be a long-period redback, this would likely mean that any mass loss from the companion would be minimal (due to the large separation between the two objects), leading to smaller eclipses, and a lower chance of multiple orbital period derivatives and/or tidal interaction. 

\subsection{Higher Order Orbital Frequency Derivatives}

Another way of checking whether the system is a redback is searching for nonzero higher orbital frequency derivatives, which are often present in redback systems. We used PINT \citep{pint} to fit the NG 15-year data for J2043+1711 with up to four orbital frequency derivatives. None of the higher order derivatives were statistically significant ($<1\sigma$), which is further evidence that J2043+1711 is not a redback pulsar. 

\subsection{Scale of Inferred Accelerations for Redbacks}

While it is not clear whether the system is actually a redback, the question that remains is that whether J2043+1711 being a redback could actually lead to the observed peculiar acceleration. To test this, we collected a number of measured $\dot{P}_b/P_b$ for redback pulsars, which are provided in Table \ref{tab:redbacks}; these quantities are the fractional change in the orbital period over time, and are related to an observed acceleration by a factor of $c$. 

The data from this table comes from the Australia Telescope National Facility Pulsar Catalogue \citep{Manchester2005}, where we have selected a sample of redback pulsars with the following properties: \begin{enumerate}
    \item $P_s<5$ ms
    \item $\dot{P_b}$ exists, and
    \item The orbital companion is a main-sequence star.
\end{enumerate} Note that for many of these systems, the orbital frequency and several time derivatives of the orbital frequency were fit to the timing data instead of the orbital period; in this case, we report \begin{equation}
    \frac{\dot{P}_b}{P_b} = -\frac{\dot{f}_b}{f_b},
\end{equation} where $f_b$ is the orbital frequency and $\dot{f}_b$ is its first time derivative. 

All of the redback pulsars have observed values of $\dot{P}_b/P_b$ that are several orders of magnitude larger than the observed value of $\dot{P}_b/P_b$ for J2043+1711. It is therefore unrealistic that the acceleration of J2043+1711 is the result of the system being a bona-fide redback pulsar, because the accelerations that would be inferred due to the orbital interactions of redbacks are hundreds to thousands of times larger than the observed peculiar acceleration of J2043+1711. 

\subsection{Core Mass--Orbital Period Relation}

Finally, redbacks of a given mass have a wide range of companion masses. However, companion masses in pulsar--Helium white dwarf systems tend to follow a narrow trend, which is known as the ``core mass--orbital period'' relation \citep{TaurisSavonije1999,Istrate2014}. If J2043+1711 is really in a bound orbit with a Helium white dwarf, the inferred mass of the companion should lie on this relation. The relations given by \citep{TaurisSavonije1999} require an estimate of the chemical composition of the companion (i.e. whether its progenitor was a population I or II star), which we do not have. However, assuming the companion's progenitor was a population I star and plugging in the orbital period for the J2043+1711 system, we predict a companion mass of 0.201 M$_\odot$; for a population II object, the companion mass is predicted to be 0.178 M$_\odot$. These values bracket the observed mass of the orbital companion, which is further evidence that the companion of J2043+1711 is in fact a Helium white dwarf, and the system is not a redback pulsar. 

Since the system does not appear to be a redback pulsar, we move forwards with the assumption that $\dot{P}_b^{\rm Int}=0$; or, in other words, that the observed peculiar acceleration is in fact an actual acceleration caused by a gravitational effect near J2043+1711.

\section{Galactic Causes} 

The MW is known to currently be in dynamical disequilibrium. This includes corrugations \citep{Xu2015}, vertical density asymmetries \citep{Widrow2012}, and phase space structures \citep{Antoja2018} in the Galactic disk (where J2043+1711 is located) that are potentially related to interactions between the MW and orbiting satellite dwarf galaxies. Notably, the acceleration field of the MW as measured using pulsars has been shown to be in substantial disequilibrium \citep{Chakrabarti2021,Donlon2024}. These disequilibrium features have been shown to either be associated with or strongly dependent on the motions of orbiting satellite dwarf galaxies \citep[e.g.][]{Gomez2017,Antoja2018}. As such, it is plausible that the observed peculiar acceleration of J2043+1711 might be due to some disequilibrium feature in the MW disk that is related to interactions between the MW and its satellite galaxies. 

Given a gravitational potential field $\Phi(\mathbf{x})$, the line-of-sight acceleration due to gravity can be calculated as $a_\mathrm{los} = -\nabla_{\hat{\mathbf{x}}_\mathrm{los}}\Phi(\mathbf{x})$, where $\nabla_{\hat{\mathbf{x}}_\mathrm{los}}$ indicates the directional derivative along our line of sight. This allows us to rewrite Equation \ref{eq:acc_breakdown} as a decomposition of the total potential at the location of PSR J2043+1711, \begin{equation}
    \nabla_{\hat{\mathbf{x}}_\mathrm{los}}\Phi_\mathrm{tot} = \nabla_{\hat{\mathbf{x}}_\mathrm{los}}\left(\Phi_\mathrm{smooth} + \Phi_\mathrm{diseq}\right),
\end{equation} where $\Phi_\mathrm{smooth}$ is the gravitational potential of the MW if it were in dynamical equilibrium (i.e., its distribution function were static), and $\Phi_\mathrm{diseq}$ is some local perturbation to the static gravitational potential field due to a disequilibrium process. Here, $\Phi_\mathrm{smooth}$ corresponds to $a_\mathrm{los}^\mathrm{Gal}$ in Equation \ref{eq:acc_breakdown}, while $\Phi_\mathrm{diseq}$ corresponds to $a_\mathrm{los}^\mathrm{Pec}$, although we use different terminology in this case because both $\Phi_\mathrm{smooth}$ and $\Phi_\mathrm{diseq}$ arise from the total gravitational potential of the Galaxy.

Since kinematic models assume dynamical equilibrium, the acceleration predicted by these models corresponds to the acceleration due to $\Phi_\mathrm{smooth}$. It is unlikely that the observed peculiar acceleration is caused by using an incorrect smooth potential model due to the observed acceleration of J2043+1711 having the opposite sign as the smooth potential model prediction. As a result, in this picture any observed deviation in the acceleration of PSR J2043+1711 must come from $\Phi_\mathrm{diseq}$.

\begin{figure}
    \centering
    \includegraphics[width=0.45\textwidth]{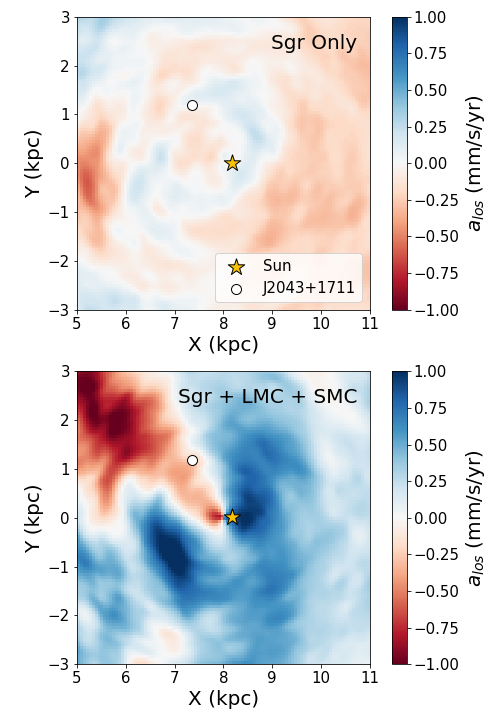}
    \caption{Simulated line-of-sight accelerations that are imparted on the Galactic disk due to interactions with satellite galaxies \citep{Chakrabarti2019}. These accelerations are for objects located in the Galactic plane, i.e. at $Z=0$. The top panel shows a simulation that only included the effects of the Sgr Dwarf Galaxy, whereas the bottom panel shows the effects of Sgr as well as the Magellanic Clouds. Positive (blue) regions are accelerating away from the Sun, and negative (red) regions are accelerating towards the Sun. These acceleration residuals were obtained by subtracting the accelerations of the initial conditions of the simulation from the accelerations at the present day; this is necessary because the potential of the smooth, unperturbed disk dominates the observed accelerations, and is unrelated to satellite interactions. The typical amplitude of peculiar accelerations due to this particular disequilibrium feature are on the scale of 0.5-1 mm/s/yr, which is not large enough to explain the observed peculiar acceleration for J2043+1711.}
    \label{fig:diseq_accel}
\end{figure}

Figure \ref{fig:diseq_accel} shows the peculiar acceleration field that is imparted on the Galactic disk plane in simulations of the Sagittarius Dwarf Galaxy and the Magellanic Clouds interacting with the MW \citep[the details of the simulations are described in][]{Chakrabarti2019}. The large mass of the Magellanic Clouds \citep[e.g.][]{Erkal2019} and the proximity of the relatively massive Sgr Dwarf Galaxy makes them the main source of external tidal forces on the MW. The perturbations caused by the satellite interaction are on the order of roughly 1 mm/s/yr near the Sun, which is too small to cause the observed peculiar acceleration of J2043+1711. 

These simulated accelerations are also interesting from a Galactic structure perspective. This $\sim1$ mm/s/yr perturbation is consistent with the findings of \cite{Donlon2024}, who noted that the vertical acceleration profile of the disk is offset by roughly that amount compared to a static equilibrium potential profile. It is worth pointing out that the line-of-sight acceleration at the location of J2043+1711 differs by less than 1 mm/s/yr between most of the disequilibrium potential profiles examined by \cite{Donlon2024} and the static equilibrium potential. Two of the \cite{Donlon2024} models (``local expansion'' and ``$\alpha$-$\beta$+2 Point Mass'') deviate from the static equilibrium potential by 3 mm/s/yr at the location of J2043+1711; however, if J2043+1711 is removed from the dataset and these models are then fit again to the remaining pulsars, these deviations become less than 1 mm/s/yr. This indicates that J2043+1711 was strongly biasing these fits, and as a result should probably be removed from future analyses of the Galactic acceleration field. 

While the actual masses of the dwarf galaxies could be larger than what was used in these simulations, which would increase the magnitude of the perturbations to the MW disk, it is unlikely that the perturbations from the dwarf in reality are an order of magnitude larger that in this simulation, which would be required to explain J2043+1711's observed peculiar acceleration. 

The presence of disequilibrium effects could reduce the significance of the peculiar acceleration on a $\sim 1\sigma$ level by altering the potential at the location of J2043+1711 in a way that partially explains the observed peculiar acceleration. However, the perturbation could just as easily work in the other direction, making the peculiar acceleration even less consistent with the effect of the underlying Galactic potential. Currently, published simulated models cannot exactly recover the observed perturbations of the Milky Way disk due to the interactions with orbiting satellites \citep{BennettBovy2021}; therefore, we urge the reader not to read into the exact shape and orientation of the acceleration profiles in Figure \ref{fig:diseq_accel}, but instead to look at the broad patterns and magnitudes of the effects, which are more likely to be representative of actual MW disk structure. 

It is also possible that dark matter substructure on $\sim$kpc scales, such as $\Lambda$CDM subhalos \citep{Diemand2007,Bullock2017} or fuzzy dark matter density fluctuations \citep{Hu2000,Hui2017} could be responsible for the anomalous acceleration. However, dark matter substructure and interactions with orbiting satellite galaxies should perturb the entire acceleration field on $\sim$kpc scales, not only a small region, and therefore cause similar correlated anomalous accelerations for other pulsars in our sample. Since we do not observe similar peculiar accelerations in any of the 11 other pulsars in the \cite{Donlon2024} dataset that have similar measurement precision as J2043+1711 ($S/N > 1$), we conclude that the peculiar acceleration is almost certainly caused by an effect that is local to J2043+1711. 


\section{Stellar Flyby}

\begin{figure}
    \centering
    \includegraphics[width=0.47\textwidth]{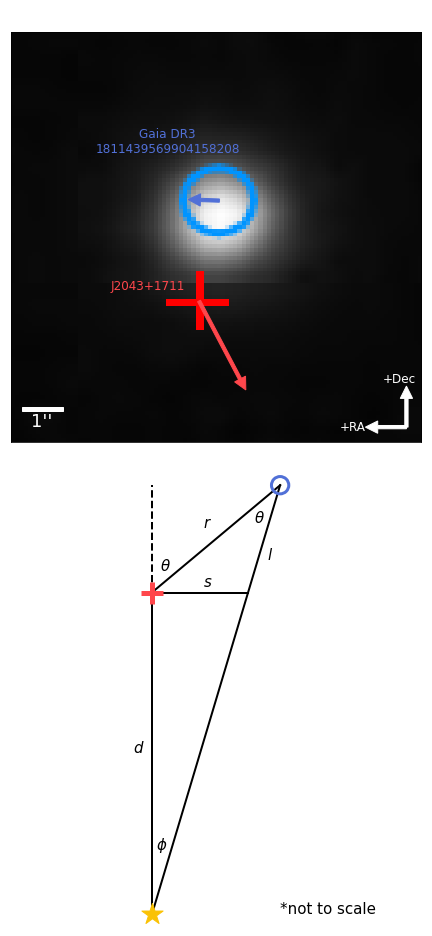}
    \caption{Configuration of J2043+1711 and \textit{Gaia} DR3 1811439569904158208. The top panel shows PS1 photometry; the blue circle shows the location of the main sequence star in \textit{Gaia} DR3, and the red cross is the location of J2043+1711. The arrows show the directions and relative magnitudes of the proper motions of each object. The bottom panel shows a diagram containing the relative distances to each object, as defined in Equation \ref{eq:flyby_config}. The golden star is the location of the Sun. Note that the two angles labeled $\theta$ are equivalent since $\phi$ is small and $l \ll d$.}
    \label{fig:photometry}
\end{figure}

In order to determine whether the observed peculiar acceleration could be caused by any known optical objects that are near J2043+1711 on the sky, we queried \textit{Gaia} DR3 \citep{GaiaDR3}. The closest star on the sky to the J2043+1711 system, designated \textit{Gaia} DR3 1811439569904158208, has an angular separation of 2.4'' from J2043+1711. Further, the parallax of this star is consistent with the parallax of J2043+1711 within their respective error bars: J2043+1711 has a parallax of 0.64 $\pm$ 0.04 mas, and the main sequence star has a parallax of 0.57 $\pm$ 0.07 mas after applying the \textit{Gaia} DR3 zero-point correction \citep{Lindegren2021}. We checked the impact of the Lutz-Kelker bias \citep{LutzKelker1973} as it applies to pulsars \citep{Verbiest2012,Igoshev2016}, and found it to be irrelevant in this case. While additional uncertainty in parallaxes fit to timing data can be potentially problematic depending on how red noise is handled for the source, it is probably not a major source of error for J2043+1711, which has no detectable red noise and consistent timing parallax values throughout previous NANOGrav data sets \citep{nanograv12.5yr,nanograv15yr}. 

The proximity of this star to J2043+1711 makes it a plausible candidate for the source of the anomalous acceleration. However, there is a chance that that the main-sequence star is unrelated to the J2043+1711 system, and just happens to be located nearby on the sky due to random chance. To test this, we estimated the odds that, given a random point on the sky, we would expect to observe a star brighter than 18th magnitude within 2.5'' of the selected location. We collected all objects in \textit{Gaia} DR3 brighter than 18th magnitude in 1 square degree of the sky centered on J2043+1711. Over 1 million Monte-Carlo samples, we observe that a randomly-selected point has a 1.6\% chance of being located within 2.5'' of a G$<$18 star. If the star is also required to have a parallax within 0.2 mas of a parallax selected at random from the \textit{Gaia} DR3 parallax distribution, this drops to just 0.6\%. This does not imply that the association of the star with J2043+1711 is simply due to random chance (especially considering the agreement in the parallaxes of the objects) but it is important to consider that this could still be a specious association.

The mass of the main-sequence star is estimated to be 0.8 $M_\odot$ based on its photometry using the \textit{Gaia} astrophysical parameter estimation pipeline \citep{gaiadr3_flame}. Using this information plus the star's location on the sky, we can determine the line-of-sight distance between the star and the pulsar that produces the observed peculiar acceleration: \begin{equation} \label{eq:flyby_config}
    a^\mathrm{Pec}_\mathrm{los} = \frac{G M_\mathrm{MS}}{r^2}\cos\theta = \frac{G M_\mathrm{MS} \sqrt{r^2 - s^2}}{r^3}, 
\end{equation} where $r$ is the total distance between the two objects, $\phi$ is the angle between our line of sight and the vector from the pulsar to the star, $s = d\cos\phi$ is the projected distance between the objects, $\theta$ is the angle at J2043+1711 between our line of sight and the main-sequence star, and $M_\mathrm{MS}$ is the mass of the main sequence star. At a distance of $d=1.7$ $\pm$ 0.1 kpc (obtained from the weighted mean of the two objects' parallaxes), $s=4000 \pm 300$ AU, which sets the line-of-sight distance between the pulsar and the star to be $l=6500\pm700$ AU, and $r=7700\pm700$ AU. 

The main-sequence star was also identified in Pan-STARRS DR2 \citep{panstarrs}, where it has the identifier PSO J310.8369+17.1921. It has a similar apparent magnitude to the \textit{Gaia} observation of the star; the Pan-STARRS photometric data for the star are given in Table \ref{tab:properties}, and are consistent with the mass and effective temperature listed for this star. This was the closest Pan-STARRS optical source to J2043+1711 on the sky. Another faint object is located 3.2'' from J2043+1711 in Pan-STARRS, which could potentially be a very low mass star near J2043+1711 given its apparent magnitude $g\sim$ 25 and color $g-i\sim2$. However, this faint object only has 2 detections in Pan-STARRS and fairly large photometric uncertainties, so it is unclear whether it is an actual star; regardless, even if it is a real star located at the optimal line-of-sight distance to maximize an imparted line-of-sight acceleration, its on-sky separation from J2043+1711 and mass of the faint object (based on its color) would not be able to explain the observed peculiar acceleration.

However, J2043+1711 and the \textit{Gaia} main-sequence star do not have similar proper motions, as shown in Table \ref{tab:properties}. This indicates that if the anomalous acceleration were indeed caused by the gravity of the main sequence star, then J2043+1711 is currently experiencing a stellar flyby. 

Renormalized Unit Weight Error, or RUWE, is a measure of how well the \textit{Gaia} astrometric pipeline is able to fit observations of an object; values above 1.25 indicate that the object is likely a member of a binary due to additional movement on the sky not associated with parallax or proper motion \citep{Penoyre2022}. Our assessment that J2043+1711 and the main-sequence star are not orbiting one another is corroborated by the fact that the RUWE value for the main sequence star in \textit{Gaia} DR3 is only 0.93. Additionally, the \textit{Gaia} DR3 internal classification schema did not identify this star as a likely binary system \citep{gaiadr3_binary}. However, it should be noted that at a distance of $r=7700$ AU, the minimum orbital period of the system is roughly 500,000 years, which is much longer than the $\sim1000$ day practical period limit of the \textit{Gaia} DR3 binary classification algorithm \citep{Holl2023}.

\;\\

\subsection{Is Gaia DR3 1811439569904158208 a Runaway?}

\begin{figure}
    \centering
    \includegraphics[width=0.47\textwidth]{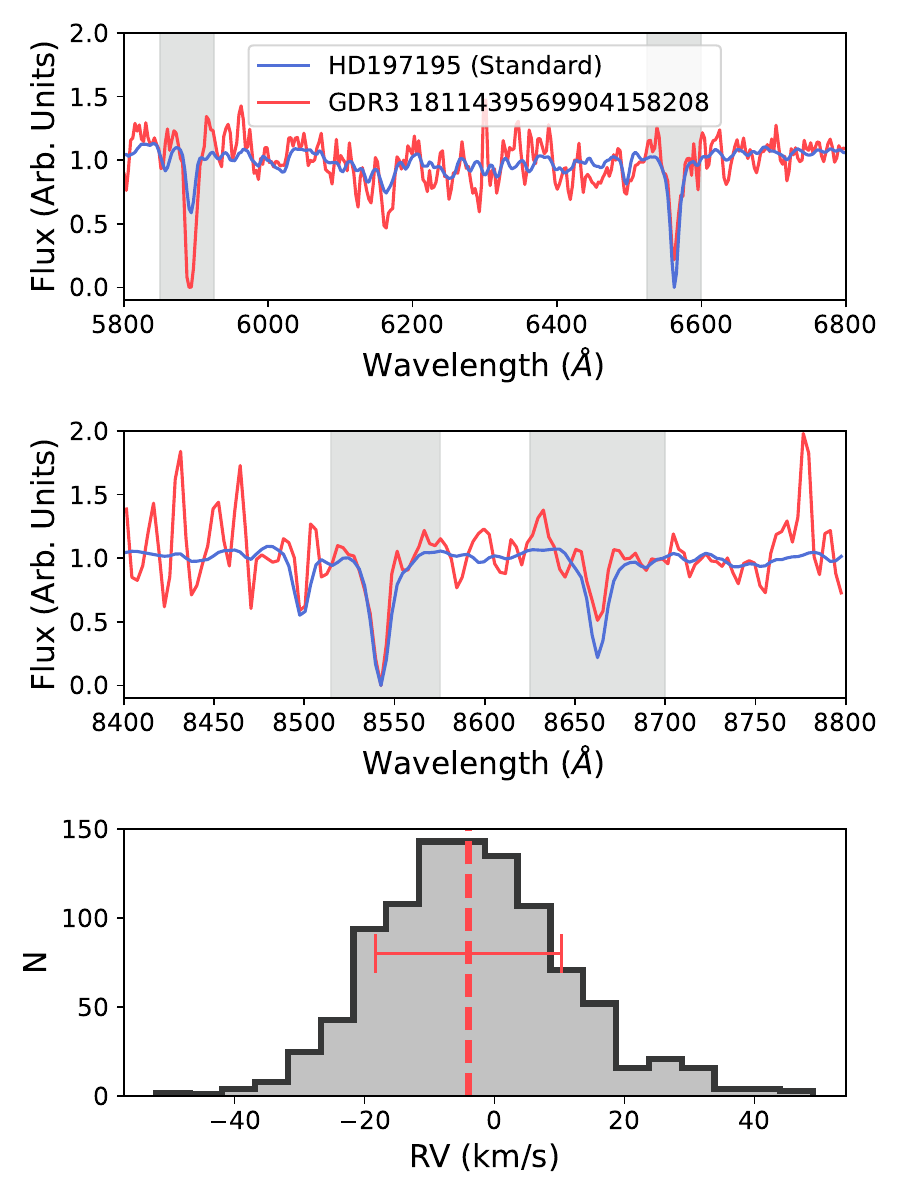}
    \caption{The spectrum and corresponding line-of-sight velocity measurement of \textit{Gaia} DR3 1811439569904158208. R$\sim$1200 spectra are shown for a standard \textit{Gaia} star with a well-known radial velocity, and \textit{Gaia} DR3 1811439569904158208. The absorption lines that were used to compute line-of-sight velocity are highlighted in gray. The distribution of line-of-sight velocity measurements (after subtracting the velocity of the standard star) from 1000 bootstraps is provided in the bottom panel; it has a median and standard deviation of RV = -4 $\pm$ 14 km/s. }
    \label{fig:spectrum}
\end{figure}

It is possible that the main-sequence star could be a runaway (or hypervelocity) star \citep{Hoogerwerf2001,Irrgang2018}; these stars are ejected from open clusters or the center of the Galaxy due to close encounters with other stars or the MW's central black hole, which gives them very large velocities of many hundreds of km/s. This could potentially explain why we observe this object so close to another star, as a runaway would have a velocity very different to that of disk stars. However, the full 3-dimensional velocity of the star is required to determine whether the main-sequence star is actually a runaway. 

Using the SuperNova Integral Field Spectrograph \citep[SNIFS,][]{SNIFS} as part of The Spectroscopic Classification of Astronomical Transients \citep[SCAT,][]{Tucker2022} Survey, we obtained 2x20 min. $R\sim1200$ spectra of \textit{Gaia} DR3 1811439569904158208 and a nearby bright \textit{Gaia} standard star (HD197195) with a known line-of-sight velocity of RV = 29 km/s \citep{Soubiran2018b}; these spectra are shown in Figure \ref{fig:spectrum}. Using the Na doublet, H$\alpha$, and 2 of the Ca triplet lines, we measure the line-of-sight velocity of \textit{Gaia} DR3 1811439569904158208 to be RV = -4 km/s. Randomly bootstrapping the observed spectra 1000 times using the uncertainty of the flux in each pixel results in an uncertainty of 14 km/s. This is likely an underestimate, as SNIFS has been shown to have a 50 km/s systematic uncertainty when measuring galaxy velocities \citep{Do2024}; however, even with this large uncertainty, the magnitude of the star's actual line-of-sight velocity should not be much larger than 50 km/s, confidently ruling out a line-of-sight velocity of hundreds of km/s. 

Combined with the proper motion of the star, we determine that the star has a 3-dimensional Galactocentric velocity of (11, 224, -16) km/s, total energy of $E\sim-1.25\times10^5$ km$^2$/s$^2$ and an angular momentum of $L_z\sim$ $-$1700 kpc km/s; these are typical values for a disk star \citep{DeasonBelokurov2024}, indicating that the main-sequence star near J2043+1711 is not a runaway. Varying the line-of-sight velocity of the star within $\pm$50 km/s does not dramatically change these results.

\subsection{Feasibility of a Flyby}

As a sanity check, we calculated how far the main sequence star would have traveled relative to J2043+1711 over the 9.2 years that NANOGrav has observed this pulsar. Assuming a relative velocity of 100 km/s (this is slightly larger than the tangential velocity inferred from the difference in the proper motions of the two objects, and is reasonable, albeit large, for a relative motion between disk objects), the main sequence star would have traveled roughly 200 AU. This is smaller than the uncertainties of the distance between the objects, implying that the variation in line-of-sight acceleration due to the star's motion over the observation baseline is smaller than the error bars on the observed line-of-sight acceleration. As such, the observed peculiar acceleration is consistent with J2043+1711 experiencing a stellar flyby. 

Similarly, we can estimate the probability of a close encounter by assuming that the Galactic midplane has uniform stellar density $\rho_*$. Then, for any given object, the number of stars within a distance $r_0$ of that object is \begin{equation}
    N(r < r_0) = \frac{4\pi}{3} \rho_* r_0^3. 
\end{equation} If we set $\rho_*$ = 0.0468 $M_\odot$/pc$^3$ \citep{Guo2020}, each object is expected to have 2.2$\times$10$^{-5}$ $M_\odot$ worth of stars within $r_0=10,000$ AU. Assuming each star is roughly 1 solar mass, we would expect to observe 1 star within $r_0$ if we observed roughly 45,000 pulsars. Considering the \cite{Donlon2024} sample of 26 pulsars, this translates to a 1 in roughly 1750 chance, indicating that it is very unexpected but not outside the realm of possibility that we observe a stellar flyby around at least one of these 26 pulsars. Note that J2043+1711 is located 0.4 kpc below the Galactic midplane, which is roughly the scale height of the thin disk; as a result, the density near J2043+1711 will be about a factor of $e$ lower than the midplane density. Similar reductions in density near other pulsars makes this assessment an overestimate, although the bias is not likely to be severe, because most pulsars are located close to the plane. 

\section{Long-Period Orbital Companion} \label{sec:long_period_companion}

While a stellar flyby is a compelling explanation for the observed peculiar acceleration of J2043+1711, it is not the only possibility. A circumbinary companion with an orbital period much longer than the NANOGrav observation baseline, i.e. a hierarchical triple, would appear to be a constant acceleration in the timing residuals for J2043+1711. This scenario could possibly cause the observed peculiar acceleration. 

\textit{Gaia} has a limiting magnitude of $G\sim21$; at a distance of 1.6 kpc, this corresponds to an absolute magnitude of $M_G\sim10$, and any objects fainter than this would not be present in the \textit{Gaia} DR3 catalog. Pan-STARRS has a lower 98\% limiting magnitude of $g_\mathrm{PS}\sim23$, requiring that any unseen object near J2043+1711 must be fainter than $M_{g_\mathrm{PS}}\sim12$. This limits any possible circumbinary companion to a neutron star, white dwarf, low-mass (sub-)stellar object, or a black hole.

\subsection{Orbital Constraints}

\begin{figure*}
    \centering
    \includegraphics[width=\textwidth]{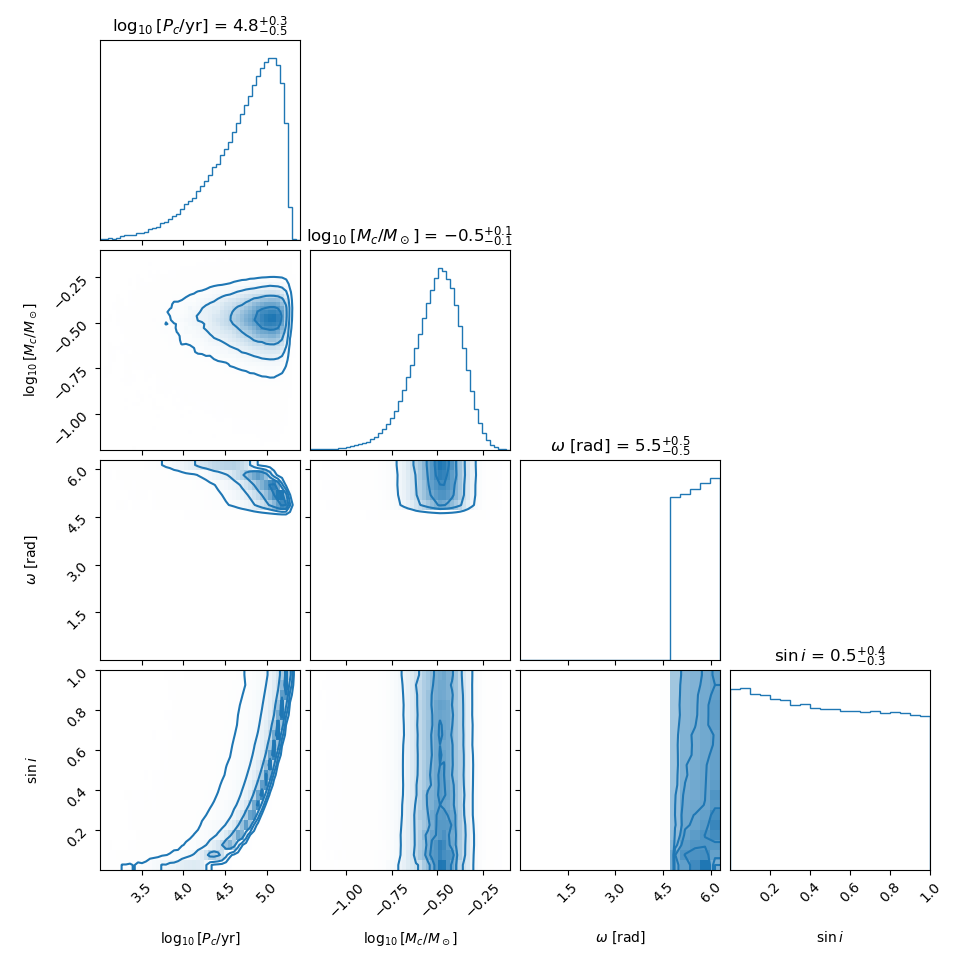}
    \caption{Posterior distributions for orbital parameters of a possible circumbinary companion around J2043+1711. The orbital period and companion mass are well constrained by the model, although the configuration of the orbit can only be constrained to lie within a specific quadrant. The inclination angle of the orbit is not constrained, and is strongly covariant with the orbital period.  }
    \label{fig:orbit_mcmc}
\end{figure*}

The pulses of a pulsar in a binary orbit will be affected by a R$\ddot{\mathrm{o}}$mer delay, which induces a sine wave in the observed times of arrival (TOAs) for that pulsar due to the motion of the pulsar around the common center of mass \citep{Lange2001}. For a circular orbit, this delay is described by \begin{equation}
    \Delta_R = x \sin i  \sin\left(\frac{2\pi}{P_c}(t - t_0) + \omega\right),
\end{equation} where $x$ is the semi-major axis of the orbit, $i$ is the orbital inclination of the system, $P_c$ is the orbital period of the outer binary, and $\omega$ is the phase of the orbit at time $t_0$. 

If the orbital period is sufficiently long compared to the observation baseline, this effect will appear to be stationary, and can be constrained through variations in the spindown rate of the pulsar \citep{JoshiRasio1997,Jones2023}. The derivatives of $\Delta_R$ with respect to time provide the line-of-sight velocity of the pulsar, acceleration, jerk, and so on. Because these quantities are related to the Doppler shift by \begin{equation}
    \frac{f^{(1)}}{f} = -\frac{a}{c},
\end{equation} where we have used the notation $f^{(n)} \equiv \dv*[n]{f}{t}$, one can continue taking time derivatives of both sides to obtain formulae for the higher-order spindown derivatives. As $\omega$ is effectively a nuisance parameter, these quantities can be evaluated at $t=t_0=0$ to obtain the general form for this system of equations \citep{Jones2023}: \begin{equation} \label{eq:jones}
    \frac{f^{(n)}}{f} = \frac{x\sin i}{c}\left(\frac{2\pi}{P_c}\right)^{n+1}(-1)^{n+1}\cos\left(\omega- \frac{n\pi}{2}\right).
\end{equation} Note that the above equation only describes a circular orbit; \cite{Jones2023} point out that this is not necessarily a good assumption, and provide instructions for how to extend their result to eccentric orbits. 

\cite{Jones2023} also note that successive spindown derivatives that are actually caused by orbital companions should follow a alternating sign pattern based on this formula, caused by successive derivatives of $\cos\omega$. As the NANOGrav 15-year fit only includes $f$ and $f^{(1)}$, we used the PINT software \citep{pint} to fit additional spindown derivatives up to $f^{(9)}$ to the J2043+1711 TOAs. The existing parameters from the NANOGrav 15-year release were also allowed to vary in this fit, but the optimal values for these parameters did not noticeably change after the addition of the new spindown derivatives.


At some point long-term variations (corresponding to higher-order spindown derivatives) in the TOAs are expected to become dominated by red noise due to effects such as interstellar space weather and the gravitational wave background \citep{nanograv15noise}. To test this, we used PINT to simulate TOAs for a pulsar with all $f^{(n)}=0$, white noise on the order of 0.1 $\mu$s (comparable to the observed noise level for J2043+1711; \citealt{nanograv15yr}), and red noise generated from a power law with spectral index $\gamma = -3.5$. Fitting the simulated residuals with only spindown derivatives produced apparently significant values with the same alternating sign pattern as is expected from Equation \ref{eq:jones}. As a result, it is difficult to claim that the higher order spindown derivatives we measure from J2043+1711 are actually due to a circumbinary companion rather than red noise. However, because the NANOGrav 15-year data set did not show significant red noise in J2043+1711 in the first place, we move forward with the assumption that each $f^{(n)}$ is actually due to an orbital companion. 

The system of equations defined by Equation \ref{eq:jones} are degenerate between $x \sin i$ and $P_c$ (a given orbital period sets the semi-major axis), so we use the measured peculiar acceleration of the system to provide a further constraint on the orbital parameters of the system in order to break this degeneracy. The line-of-sight acceleration is given as \begin{equation} \label{eq:alos_constraint}
    a_\mathrm{los}^\mathrm{Pec} = \frac{G M_c}{x^2}\sin i \sin(-\omega),
\end{equation} where $x = x(P_c, M_c)$ can be determined from Kepler's Third Law. 

However, note that Equation \ref{eq:alos_constraint} uses the line-of-sight acceleration, which is proportional to $f^{(1)}/f$; this means that Equations \ref{eq:jones} and \ref{eq:alos_constraint} are not independent. Although adding this constraint allows us to break the degeneracy between $x$ and $\sin i$, the inclination angle cannot be fit and remains a free parameter.

\subsection{Markov Chain Monte Carlo Results}

Using these constraints, we set up a Markov Chain Monte Carlo (MCMC) simulation using the \textbf{emcee} software \citep{emcee} in order to obtain constraints on the possible orbital parameters for a circumbinary orbital companion. This procedure is similar to that of \cite{Kaplan2016} and \cite{Bassa2016}, who used comparable fitting methods to discover an orbiting companion around J1024$-$0719 based on its peculiar acceleration.

The likelihood function for this simulation is defined as \begin{equation}
    \ln\mathcal{L}\left(a_\mathrm{los,obs}^\mathrm{Pec}, f^{(n)} | P_c, M_c, i, \omega\right) = \ln\mathcal{L}_{a} + \ln\mathcal{L}_{f}, 
\end{equation} where the log-likelihood for observing a given $a_\mathrm{los}^{Pec}$ is \begin{equation}
  \ln\mathcal{L}_a = - \frac{\left(a_\mathrm{los,obs}^\mathrm{Pec} - a_\mathrm{los,model}^\mathrm{Pec}\right)^2}{2\sigma_{a_\mathrm{los}^\mathrm{Pec}}^2},
\end{equation} and the likelihood for observing the measured spindown derivatives is \begin{equation}
    \ln\mathcal{L}_f = - \sum_{n=1}^3\frac{\left(f^{(n)}_\mathrm{obs} - f^{(n)}_\mathrm{model}\right)^2}{2\sigma_{f^{(n)}}^2}, 
\end{equation} where $f^{(n)}_\mathrm{model}$ and $a_\mathrm{los,model}^\mathrm{Pec}$ are obtained from Equations \ref{eq:jones} and \ref{eq:alos_constraint}, respectively. 

Note that only the first three spindown derivatives are used in this setup. This is because higher-order derivatives have progressively larger uncertainties; in order to constrain the three free orbital parameters (as $\sin i$ cannot be constrained), we require three constraints, which are provided by the first three spindown derivatives. Using all nine spindown derivatives does not substantially change the median values of the posterior distributions, but it does increase the uncertainties on the reported values. 

The results of the MCMC simulation are shown in Figure \ref{fig:orbit_mcmc}. Although the orbital period and companion mass are well constrained, although $\omega$ is only constrained to lie within a specific quadrant. The best-fit parameters of the MCMC code produce to an orbit with $P_c=80$ kyr and $M_c=0.3$ $M_\odot$, corresponding to a semi-major axis of $x\approx2000$ AU. This mass suggests that the companion would probably be a white dwarf or low-mass main sequence star, which would be allowed by the constraints on absolute magnitude given by \textit{Gaia} and PS1 ($M_g\gtrsim$12). It should be noted that white-dwarf masses below 0.5 M$_\odot$ cannot be reached through isolated stellar evolution \citep[i.e.][]{El-Badry2018}; therefore, if this object is a white dwarf, it has likely interacted with at least one other object.

\subsection{Comments on the Orbital Configuration of a Possible Triple System}

If J2043+1711 does in fact have a circumbinary companion, this would make it a particularly interesting system for additional study. Pulsar triple systems evolve through complex processes, providing laboratories that test our physical understanding of supernovae and orbital dynamics. 

The formation of the inner white dwarf companion to J2043+1711 can be explained through binary evolution where the companion transferred mass onto J2043+1711, recycling the pulsar. The candidate outer companion is likely a low-mass main sequence star given its estimated mass of $0.31^{+0.48}_{-0.30}$ $M_\odot$, although we cannot rule out the candidate companion being a white dwarf. As such, there are several possibilities for the formation of an outer companion, including (but not limited to):  \begin{enumerate}
    \item The outer companion evolved mostly in isolation. It would be difficult to keep such an object bound to the system after a supernova, although there is some evidence for pulsar formation without large natal kicks \citep{Gomez2018,Willcox2021}.
    \item The outer companion evolved elsewhere and was then captured by the inner binary, potentially through an interaction between the inner J2043+1711 system and a second binary system \citep{Mardling2001}.
    \item The current outer companion interacted with the pulsar, could potentially undergo mass transfer, and then swapped positions with the current inner white dwarf companion without the triple becoming unstable. This scenario is sensitive to the masses, eccentricity, and inclinations of the orbits in the system, and often leads to a stable configuration where $a_\mathrm{out}/a_\mathrm{in}\sim3$ \citep{Mardling2001}, although subsequent supernova kicks could easily change the semi-major axes of the system. 
    \item Mass loss in the inner binary increased the semi-major axis of the orbit of the outer companion. This could happen either adiabatically or instantaneously through a supernova; the first scenario would imply a low orbital eccentricity for the outer companion, whereas the second scenario implies a large eccentricity (although the eccentricity of the outer system could possibly continue to evolve over time due to other processes).
\end{enumerate}

Particularly relevant for evolved triple systems is the Kozai-Lidov Effect \citep{ShappeeThompson2013,Naoz2016}, where interactions between the inner binary and a tertiary companion lead to periodic oscillations in eccentricity and inclination of the inner binary. This allows for the inner binary to reach essentially arbitrarily circular eccentricities ($e\lesssim10^{-5}$) and could cause tidal interactions or mass transfer in the inner binary, which would potentially explain the current recycled state of J2043+1711. The eccentricity of the inner binary and estimated semi-major axis of $\sim2000$ AU ($a_{\rm out}/a_{\rm in}\sim70,000$) for the outer orbit are realistic for this scenario. 

It is difficult to make any definitive statement at this point given our relatively weak constraints on the J2043+1711 system. Further complications such as uncertainties in the workings of natal kicks and recycled pulsar formation mechanisms make this a difficult theoretical problem that depends on poorly constrained assumptions. It initially seems unlikely that the outer companion would remain bound to the system, which leads one towards a scenario where the outer companion interacted in some way with the inner binary leading to the present configuration of the system; however, this is speculative, and a series of simulations would be required to understand the relative probabilities of the above scenarios, as well as identifying any other possibilities. 

\section{Discussion}

\subsection{Non-trivial Tidal Effects}

Since spider pulsars are believed to be a transitional stage of evolution between a classical and millisecond pulsars, one supposes that there may be a phase of the pulsar's evolution where it is no longer a true redback, but weak evolutionary effects such as mass loss and tidal interactions that could affect the orbital period of the binary system are still present. For example, the companion of J2043+1711 could be a semi-degenerate extended main sequence star, or a Helium white dwarf that still has an extended hydrogen shell. 

Classically, tidal effects will cause the orbit to decay, resulting in an observed negative $\dot{P}_b$ \citep{Chakrabarti2022}. This can be calculated as \begin{equation} \label{eq:tidal}
    \dot{P}_b^{\rm Tidal} = -\frac{27 \pi}{2} \sum_{i=1,2} \frac{1}{Q_i'}\frac{M_j}{M_i}\left(\frac{R_i}{a}\right)^5,
\end{equation} where $j\neq i$, $R_i$ is the radius of each object, $M_i$ is the mass of each object, $a$ is the semimajor axis of the orbit, and $Q_i'$ is the reduced tidal quality factor for an object, which depends on the internal properties and distribution of a body. A rough value of $Q_i'$ for an extended white dwarf is $10^{11}$ \citep{Fuller2013}, and $Q_i'$ for a low-mass main sequence star is something like 10$^8$ \citep{Mathis2015}; a neutron star will be much more rigid than these examples, so we let $1/{Q_i'}\rightarrow0$ for the pulsar. Additional tidal terms are typically included to account for $\dot{P}_b$ due to apsidal precession in eccentric orbits; however, this effect is negligible for J2043+1711 given its small eccentricity. 

Note that this value is always negative, and the observed $\dot{P}_b$ for J2043+1711 is positive; this indicates that if the observed peculiar acceleration is in-fact due to tidal interactions, they must be complicated. It is clear that the known interactions in redbacks is already more complicated than this simplified picture, as the pulsars in Table \ref{tab:redbacks} have both positive and negative intrinsic accelerations. Because the actual processes at play are unknown, we calculate only the rough order of magnitude for this type of effect below, as the reader should note that the physics of such a phenomenon are not well understood, nor are they handled in a robust way in this work. 

In order to obtain the observed change in the orbital period for J2043+1711 ($\dot{P}_b\sim5.7\times10^{-14}$ s/s), Equation \ref{eq:tidal} requires that a low-mass main sequence star needs a radius of 0.2 R$_\odot$, and a white dwarf companion requires a radius of 0.8 R$_\odot$. This is an unreasonable radius for a white dwarf, but possible for an extended low-mass main sequence companion. If the companion had this radius and an effective temperature of 4000 K (a typical temperature for a K-M type star, as these extended main sequence companions have normal spectral types and therefore normal temperatures), then it would have an absolute magnitude of $M_g\sim9$. This type of objects should have been seen in either the \textit{Gaia} or PS1 catalogs, which we have determined above should see objects as faint as $M_g\sim11-12$ at the distance of J2043+1711. However, it is difficult to know exactly what $Q_i'$ is for an exotic object such as an extended low-mass main sequence star, which could easily change this final result by a magnitude or more. So, while it seems unlikely based on this preliminary analysis that J2043+1711 is experiencing non-trivial tidal interaction effects as it transitions from a bona-fide redback to a non-spider pulsar, we cannot entirely rule out this possibility.

\subsection{Negative Dynamical Friction}

Massive objects moving through dense space produce an overdense wake behind their direction of motion, which imparts an acceleration antiparallel to the object's motion vector \citep[see for example][]{Chandrasekhar1943}. This process is known as dynamical friction, and in the presence of a weak outflow from the pulsar, can be calculated as\begin{equation}
    a_\mathrm{DF} = 0.275 G V_p \sqrt{\frac{\dot{E}}{c} \frac{\rho}{V_w}},
\end{equation} where $\dot{E}$ is the spindown luminosity of the pulsar, $\rho$ is the density of the interstellar medium (ISM, roughly 1 atom per cubic centimeter), $V_w$ is the velocity of the pulsar's wind emission, and $V_p$ is the velocity of the pulsar relative to the ISM \citep{Gruzinov2020}. For a millisecond pulsar traveling at 75 km/s, this effect is extremely small, roughly 1$\times10^{-14}$ mm/s/yr.

The jets emitted by a pulsar carve out a cavity in the ISM, surrounded by an overdense bow shock where the pressure generated by the pulsar emission is equal to the ram pressure of the ISM \citep[e.g.][]{Ramirez-Ruiz2019}. This bow shock is asymmetric, being closest to the pulsar in the direction of the pulsar's motion. The resulting overdensity in the direction of motion (and lack of nearby material antiparallel to the motion vector) can impart an acceleration onto the pulsar, leading to a negative dynamical friction of the system \citep{Li2020,Gruzinov2020}. The negative dynamical friction of the system is calculated as \begin{equation}
    a_\mathrm{NDF} = 2.31 \frac{G}{V_p} \sqrt{\frac{\dot{E}}{c}\rho}. 
\end{equation} For a typical millisecond pulsar, $a_\mathrm{NDF}\sim1\times10^{-6}$ mm/s/yr. This effect is far too small to be observed in current pulsar acceleration studies. As a result, peculiarities in the bow shock or emission of J2043+1711 cannot cause its observed peculiar acceleration.

\subsection{Comparing the Strengths of Various Accelerations}

\begin{table}[]
    \centering 
    \begin{tabular}{lr} \hline \hline
        Source & Strength \\ 
         & \textit{(mm/s/yr)} \\ \hline
        *Galactic Acceleration & 2 -- 10 \\
        *Galactic Disk Disequilibria & 0.5 -- 1 \\
        1 kpc from the LMC (1.4$\times10^{11}$ M$_\odot$) & 9 \\
        10 kpc from the LMC & 1 \\ 
        *60 kpc from the LMC & 0.2 \\
        2 kpc from a $10^{9}$ M$_\odot$ dwarf galaxy & 4 \\ 
        1 pc from a $10^{6}$ M$_\odot$ globular cluster & 2000 \\
        10 pc from a $10^{6}$ M$_\odot$ globular cluster & 50 \\
        1 kpc from a $10^8$ M$_\odot$ cold dark matter subhalo & 0.5 \\
        200 pc from a $10^7$ M$_\odot$ giant molecular cloud & 1 \\
        5 pc from a $10^4$ M$_\odot$ giant molecular cloud & 2 \\
        1 pc from a 1 M$_\odot$ star & 0.005 \\
        10,000 AU from a 1 M$_\odot$ star & 2 \\
        *2,000 AU from a 0.3 M$_\odot$ star & 14 \\
        100 AU from a 1 M$_\odot$ star & 20,000 \\
        30 AU from a Neptune-mass planet & 10 \\
        5 AU from a Jupiter-mass planet & 7000 \\ 
        1 AU from an Earth-mass planet & 600 \\
        0.4 AU from a Mercury-mass planet & 200 \\
        2.5 AU from a 10$^{19}$ kg asteroid & 0.0002 \\
        *Negative dynamical friction & 0.000\,001 \\
        Inferred Redback Pulsar Acceleration & $\gtrsim$10,000 \\
        \hline
    \end{tabular}
    \caption{Potential sources of acceleration on objects in the MW, and the approximate strength of each effect. Items with an asterisk are scenarios which potentially affect the J2043+1711 system.}
    \label{tab:accels}
\end{table}

There are many different effects that could potentially cause the acceleration of a given object in the Galaxy. These accelerations can point in different directions, and the strength of each effect spans a wide range of orders of magnitude. In Table \ref{tab:accels} we provide a list of possible sources of accelerations on objects in the MW, and the approximate magnitude of each effect.

\section{Conclusions} \label{sec:conclusion}

\begin{figure}
    \centering
    \includegraphics[width=0.45\textwidth]{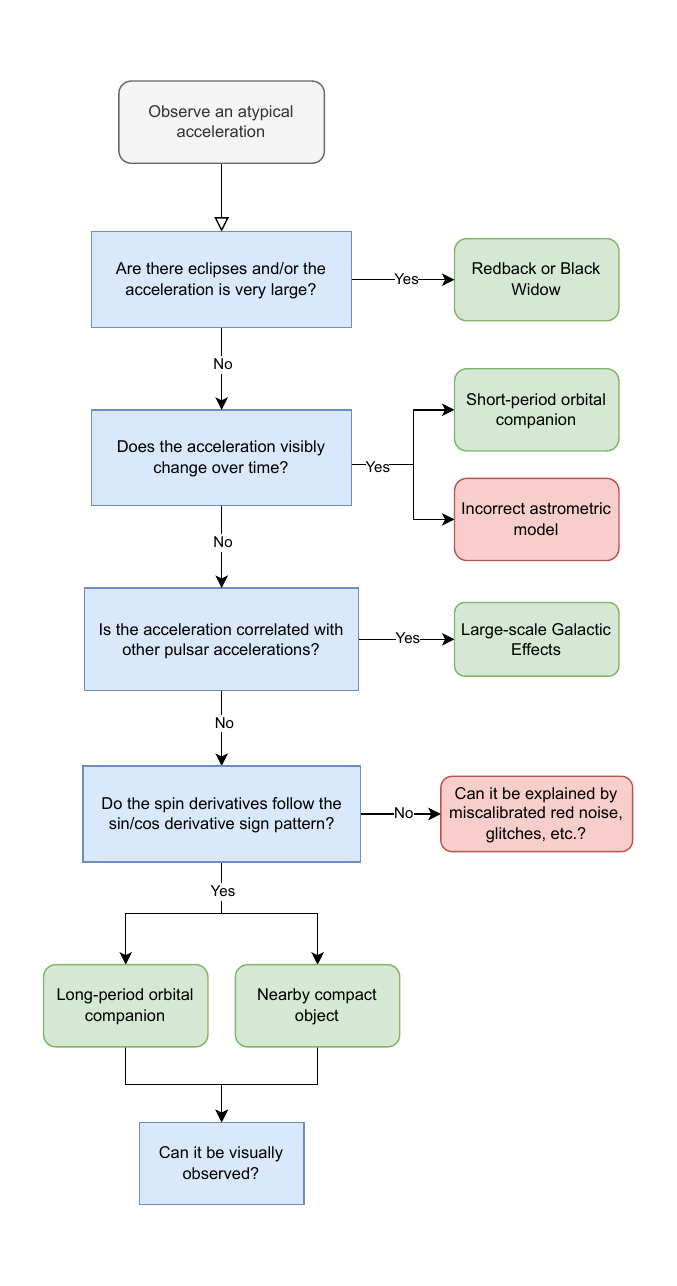}
    \caption{A guide for identifying the cause of anomalous accelerations in pulsars. }
    \label{fig:flowchart}
\end{figure}

We have shown that the binary millisecond pulsar J2043+1711 has a substantial peculiar acceleration as measured from the rate-of-change of its binary orbital period. The magnitude of the peculiar acceleration is 3.2 $\pm$ 0.8 mm/s/yr away from the Sun. The observed acceleration is the opposite sign than the predicted acceleration due to the Galaxy, and is a 4$\sigma$ deviation from the values predicted by equilibrium Milky Way models. 

We show that the magnitude of the peculiar acceleration is too large to be explained by disequilibrium effects of the Milky Way interacting with orbiting dwarf galaxies. While it is possible that this acceleration is caused by (dark matter) substructure in the Galactic density field, it is unlikely that this is the case because we do not see similar peculiar accelerations in other millisecond binary pulsars. 

Similarly, if J2043+1711 were a redback pulsar, this could potentially explain an observed peculiar acceleration. However, the magnitude of intrinsic accelerations generated by redback pulsars are much larger (by several orders of magnitude) than the observed signal for J2043+1711. For this reason, as well as the fact that no eclipses are observed for the system, J2043+1711 has no higher-order orbital period derivatives, and its companion follows the core mass--orbital period relation for Helium white dwarfs, we conclude that the redback scenario cannot be the cause for the observed peculiar acceleration. While J2043+1711 is clearly not a bona-fide redback, it is possible that the system is in some intermediate stage of evolution between a redback and a typical millisecond pulsar. In this scenario, weak tidal effects from an extended low-mass main sequence star companion could potentially contribute the observed peculiar acceleration; however, this would require that the companion have an absolute magnitude of $M_g\sim9$, which should have been observed in the \textit{Gaia} and PS1 surveys. We note that the uncertainty on this magnitude could potentially be a magnitude or more, as the tidal quality factor of this type of exotic object is not well known, making it difficult to positively rule out this possibility. 

We examine two potential local causes for the anomalous acceleration; a stellar flyby, and a long-period orbital companion. A general outline for identifying the source of an anomalous acceleration is provided in Figure \ref{fig:flowchart}, which is followed by this work.  Identifying anomalous accelerations will soon become possible with other techniques to measure Galactic accelerations, including in the near future eclipse timing \citep{Chakrabarti2022}, and farther down the road, extreme-precision radial velocity observations \citep{Chakrabarti2020}, where one can expect to follow a similar procedure as outlined in Figure \ref{fig:flowchart} (albeit with different observational schemes).

The star \textit{Gaia} DR3 1811439569904158208 is located close to J2043+1711 both in distance and on-sky position. This main-sequence star has a mass of roughly 0.8 $M_\odot$, and could explain the observed peculiar acceleration due to its proximity to J2043+171. Assuming that this star is the sole cause of the peculiar acceleration, we constrain its line-of-sight distance from J2043+1711 to be 6500 $\pm$ 700 AU, and a total distance from the pulsar of 7700 $\pm$ 700 AU. However, because the star has a substantially different proper motion than the pulsar, this star cannot be gravitationally bound to J2043+1711; rather, the pulsar would be experiencing a stellar flyby. We obtained spectra for this star and measure the magnitude of its line-of-sight velocity to be $\lesssim$ 50 km/s, indicating that it cannot be a runaway (hypervelocity) star. 

We fit higher-order spindown derivatives of J2043+1711's pulse TOAs in order to constrain the properties of a possible circumbinary orbital system. We identify plausible parameters for such a system; the observed properties of J2043+1711 are consistent with the existence of a circumbinary companion that has an orbital period of $60^{+80}_{-40}$ kyr, a semi-major axis of $1900^{+1100}_{-1000}$ AU, and a mass of $0.31^{+0.08}_{-0.06}$ $M_\odot$. Any companion would likely be a white dwarf or low-mass main sequence star due to its mass and the requirement that it be faint enough to not be identified in \textit{Gaia} DR3 or Pan-STARRS DR1. 

It is possible that the observed peculiar acceleration is really a combination of multiple effects, i.e., the nearby star plus a circumbinary orbital companion contribute accelerations that when summed together equal the observed peculiar acceleration. It is possible that with future observations of J2043+1711 (which are ongoing with NANOGrav), we might be able to rule out one of these possibilities for the source of the peculiar acceleration. In particular, a longer timing baseline for J2043+1711 will improve uncertainties on higher-order spindown derivatives, which will allow for a higher degree of accuracy when estimating orbital parameters for a candidate circumbinary companion. Alternatively, optical observations down to $\sim$26th magnitude would be able to prove or rule out the existence of a tertiary low-mass stellar or white dwarf companion to the J2043+1711 system.


\acknowledgments

We would like to thank Alice Quillen and Shami Chatterjee for helpful conversations which improved this work. 

This paper includes archived data obtained through the Australia Telescope Online Archive (\url{http://atoa.atnf.csiro.au}).

The NANOGrav Collaboration receives support from the National Science Foundation (NSF) Physics Frontiers Center award No. 2020265. The Arecibo Observatory was a facility of the NSF operated under cooperative agreement (AST-1744119) by the University of Central Florida (UCF) in alliance with Universidad Ana G. Méndez (UAGM) and Yang Enterprises (YEI), Inc.

SC acknowledges support from NSF AAG 2009828 and STSCI GO award 17505. L.B. acknowledges support from the National Science Foundation under award AST-1909933 and from the Research Corporation for Science Advancement under Cottrell Scholar Award No. 27553.
P.R.B. is supported by the Science and Technology Facilities Council, grant number ST/W000946/1.
S.B. gratefully acknowledges the support of a Sloan Fellowship, and the support of NSF under award \#1815664.
M.C. and S.R.T. acknowledge support from NSF AST-2007993.
M.C. and N.S.P. were supported by the Vanderbilt Initiative in Data Intensive Astrophysics (VIDA) Fellowship.
Support for this work was provided by the NSF through the Grote Reber Fellowship Program administered by Associated Universities, Inc./National Radio Astronomy Observatory.
Pulsar research at UBC is supported by an NSERC Discovery Grant and by CIFAR.
K.C. is supported by a UBC Four Year Fellowship (6456).
M.E.D. acknowledges support from the Naval Research Laboratory by NASA under contract S-15633Y.
T.D. and M.T.L. are supported by an NSF Astronomy and Astrophysics Grant (AAG) award number 2009468.
E.C.F. is supported by NASA under award number 80GSFC21M0002.
G.E.F., S.C.S., and S.J.V. are supported by NSF award PHY-2011772.
K.A.G. and S.R.T. acknowledge support from an NSF CAREER award \#2146016.
A.D.J. and M.V. acknowledge support from the Caltech and Jet Propulsion Laboratory President's and Director's Research and Development Fund.
A.D.J. acknowledges support from the Sloan Foundation.
The work of N.La., X.S., and D.W. is partly supported by the George and Hannah Bolinger Memorial Fund in the College of Science at Oregon State University.
N.La. acknowledges the support from Larry W. Martin and Joyce B. O'Neill Endowed Fellowship in the College of Science at Oregon State University.
Part of this research was carried out at the Jet Propulsion Laboratory, California Institute of Technology, under a contract with the National Aeronautics and Space Administration (80NM0018D0004).
D.R.L. and M.A.M. are supported by NSF \#1458952.
M.A.M. is supported by NSF \#2009425.
C.M.F.M. was supported in part by the National Science Foundation under Grants No. NSF PHY-1748958 and AST-2106552.
A.Mi. is supported by the Deutsche Forschungsgemeinschaft under Germany's Excellence Strategy - EXC 2121 Quantum Universe - 390833306.
The Dunlap Institute is funded by an endowment established by the David Dunlap family and the University of Toronto.
K.D.O. was supported in part by NSF Grant No. 2207267.
T.T.P. acknowledges support from the Extragalactic Astrophysics Research Group at E\"{o}tv\"{o}s Lor\'{a}nd University, funded by the E\"{o}tv\"{o}s Lor\'{a}nd Research Network (ELKH), which was used during the development of this research.
H.A.R. is supported by NSF Partnerships for Research and Education in Physics (PREP) award No. 2216793.
S.M.R. and I.H.S. are CIFAR Fellows.
Portions of this work performed at NRL were supported by ONR 6.1 basic research funding.
J.D.R. also acknowledges support from start-up funds from Texas Tech University.
J.S. is supported by an NSF Astronomy and Astrophysics Postdoctoral Fellowship under award AST-2202388, and acknowledges previous support by the NSF under award 1847938.
C.U. acknowledges support from BGU (Kreitman fellowship), and the Council for Higher Education and Israel Academy of Sciences and Humanities (Excellence fellowship).
C.A.W. acknowledges support from CIERA, the Adler Planetarium, and the Brinson Foundation through a CIERA-Adler postdoctoral fellowship.
O.Y. is supported by the National Science Foundation Graduate Research Fellowship under Grant No. DGE-2139292.

The Pan-STARRS1 Surveys (PS1) and the PS1 public science archive have been made possible through contributions by the Institute for Astronomy, the University of Hawaii, the Pan-STARRS Project Office, the Max-Planck Society and its participating institutes, the Max Planck Institute for Astronomy, Heidelberg and the Max Planck Institute for Extraterrestrial Physics, Garching, The Johns Hopkins University, Durham University, the University of Edinburgh, the Queen's University Belfast, the Harvard-Smithsonian Center for Astrophysics, the Las Cumbres Observatory Global Telescope Network Incorporated, the National Central University of Taiwan, the Space Telescope Science Institute, the National Aeronautics and Space Administration under Grant No. NNX08AR22G issued through the Planetary Science Division of the NASA Science Mission Directorate, the National Science Foundation Grant No. AST-1238877, the University of Maryland, Eotvos Lorand University (ELTE), the Los Alamos National Laboratory, and the Gordon and Betty Moore Foundation.

\software{Numpy \citep{numpy}, Scipy \citep{scipy}, Sci-kit Learn \citep{scikit-learn}, Matplotlib \citep{matplotlib}, PINT \citep{pint}, Gala \citep{Gala}}

\clearpage

\bibliographystyle{aasjournal}
\bibliography{references.bib}

%
%
%

\appendix 

\section{Calculation of Radial Acceleration Bias for a Given Pulsar} \label{app:d24_bias}

\begin{figure}
    \centering
    \includegraphics[width=\linewidth]{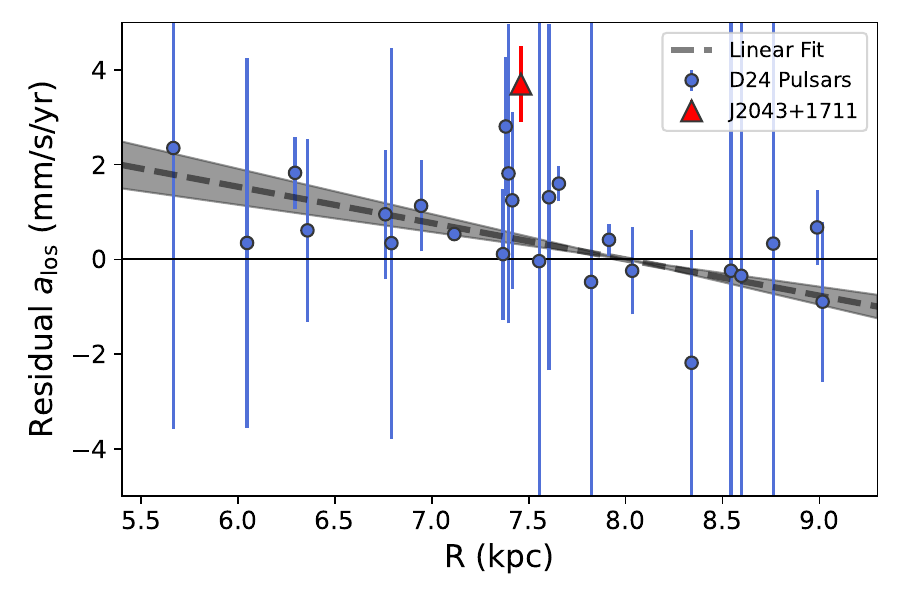}
    \caption{Line-of-sight acceleration bias in pulsars from \cite{Donlon2024}, as a function of distance from the center of the Galaxy. J2043+1711 is highlighted as a red triangle. A linear fit constrained so that it is equal to zero at the Solar location is provided as a dashed line, and the grey region indicates the 1$\sigma$ error-bars of that value. }
    \label{fig:app_bias_calc}
\end{figure}

In Section III of \cite{Donlon2024}, it is argued that pulsars near the Sun have biased accelerations. This is shown in their Figure 3, in which they plot a linear fit to the residuals of the line-of-sight accelerations for each pulsar, minus the expected Galactic acceleration for those pulsars, as a function of their distance from the Galactic center ($R$). These residuals should be distributed evenly about zero; instead, there is a clear linear trend as a function of their distance from the center of the Galaxy. It is interesting to note that \cite{Donlon2024} did not include J2043+1711 in this fit, stating that it was an outlier that substantially affected the quality of the fit. 

A reproduction of this fit is provided in Figure \ref{fig:app_bias_calc}, which was calculated using the \textbf{curve\_fit} function from the \textit{SciPy} Python package. We show the acceleration residuals of the D24 pulsars, along with a linear fit to this data. Note that here, we have constrained the linear fit so that it must be equal to zero at the position of the Sun ($R=8$ kpc). This was overlooked by \cite{Donlon2024} -- by definition, the Sun must have zero line-of-sight acceleration. This adjustment greatly reduces the uncertainty in the fit slope compared to a fit where the intercept is also allowed to vary.

The bias $a_\mathrm{los,B}$ in the line-of-sight acceleration of any given pulsar can be calculated based on its Galacticentric radius, \begin{equation}
    a_\mathrm{los,B} = m (R - 8\;\mathrm{ kpc}),
\end{equation} where our optimized value for $m$ is -0.77 $\pm$ 0.19 mm/s/yr/kpc. At the location of J2043+1711, $R$ = 7.4 kpc, leading to an acceleration bias of 0.46 $\pm$ 0.11 mm/s/yr. 

Note that including this effect \textit{decreases} the significance of the J2043+1711 peculiar acceleration. This is to say that the observed peculiar acceleration is not created by our treatment of the radial acceleration bias, and that the uncertainty associated with this bias is not large enough to significantly impact the validity of our claims.

\end{document}